\documentclass[a4paper,twoside]{article}        
\usepackage{epsfig}
\usepackage{subcaption}
\usepackage{calc}
\usepackage{amssymb}
\usepackage{amstext}
\usepackage{amsmath}
\usepackage{amsthm}
\usepackage{multicol}
\usepackage{pslatex}
\usepackage{apalike}
\usepackage[bottom]{footmisc}
\usepackage{cite}
\usepackage{amsmath,amssymb,amsfonts}
\usepackage{algorithmic}
\usepackage{graphicx}
\usepackage{textcomp}
\usepackage{xcolor}
\usepackage{acro}
\usepackage{hyperref}
\usepackage{enumitem}
\usepackage{makecell}
\usepackage{tikz}
\usetikzlibrary{positioning, fit, calc, shapes}
\usepackage{amsthm}
\newtheorem{definition}{Definition}
\usepackage{xcolor}
\usepackage{listings}
\usetikzlibrary{positioning, fit, calc, shapes}
\usepackage{SCITEPRESS}

\begin{document}

\title{Evaluating the Fork-Awareness of Coverage-Guided Fuzzers}

\author{
\authorname{Marcello Maugeri\sup{1}\orcidAuthor{0000-0002-6585-5494},
Cristian Daniele\sup{2}\orcidAuthor{0000-0001-7435-4176},
Giampaolo Bella\sup{1}\orcidAuthor{0000-0002-7615-8643}, 
and Erik Poll\sup{2}\orcidAuthor{0000-0003-4635-187X}}
\affiliation{\sup{1}Department of Maths and Computer Science, University of Catania, Catania, Italy}
\affiliation{\sup{2}Department of Digital Security, Radboud University, Nijmegen, The Netherlands}
\email{marcello.maugeri@phd.unict.it, cristian.daniele@ru.nl, giampaolo.bella@unict.it, erikpoll@cs.ru.nl}
}

\keywords{Fuzzing, Fork, Security Testing, Software Security}

\abstract{Fuzz testing (or fuzzing) is an effective technique used to find security vulnerabilities.
It consists of feeding a software under test with malformed inputs, waiting for a weird system behaviour (often a crash of the system).
Over the years, different approaches have been developed, and among the most popular lies the coverage-based one. It relies on the instrumentation of the system to generate inputs able to cover as much code as possible.
The success of this approach is also due to its usability as fuzzing techniques research approaches that do not require (or only partial require) human interactions. 
Despite the efforts, devising a fully-automated fuzzer still seems to be a challenging task.
Target systems may be very complex; they may integrate cryptographic primitives, compute and verify check-sums and employ \textit{forks} to enhance the system security, achieve better performances or \textit{manage different connections at the same time}. 
This paper introduces the \textit{fork-awareness} property to express the fuzzer ability to manage systems using forks.
This property is leveraged to evaluate 14 of the most widely coverage-guided fuzzers and highlight how current fuzzers are ineffective against systems using \textit{forks}.}

\onecolumn \maketitle \normalsize \setcounter{footnote}{0} \vfill

\section{Introduction}\label{sec:intro}

In the last years, plenty of fuzzers have been developed to deal with sophisticated software and nowadays it is extremely common that network systems employ forks to deal with different connections at the same time.
This leads to 1) the need to devise accurate and ad-hoc fuzzers and 2) the need to evaluate these fuzzer according to their ability to cope with such advanced systems.

Unfortunately, as pointed out in \cite{hazimeh2020magma}, it is not easy to benchmark all of them since the fuzzers are very different from each other.
Metzman et al. faced this problem by devising \textit{FuzzBench}\cite{metzman2021fuzzbench}, an open-source service for the evaluations of stateless fuzzers.
Later, Natella and Pham presented \textit{ProFuzzBench}\cite{natella2021profuzzbench}, which similarly to \textit{FuzzBench} provides a service to evaluate stateful fuzzers.

Although FuzzBench includes a sample of real word programs and ProFuzzBench includes different network systems (i.e. systems that often employ forks to deal with multiple connections\cite{tanenbaum2009modern}), they do not evaluate the ability of the fuzzers to cope with programs that use forks.

Despite \textit{forks} representing the only way to create a new process \cite{tanenbaum2009modern}, experimental results have shown that current fuzzers cannot deal with forked processes.

The existing approach merely relies on code modifications to remove the forks. Unfortunately, this approach goes against the willingness to reduce manual work and improve automation during a fuzzing campaign.\cite{boehme2021fuzzing}.

In this work, we explore and classify the limitations current fuzzers exhibit in front of forking programs.

In summary, this paper:
\begin{enumerate}
    \item devises a novel property capturing the ability of fuzzers to deal with forks appropriately;
    \item evaluates 14 coverage-guided fuzzers based on this property;
    \item proposes possible improvements to the current state-of-the-art and future directions.
\end{enumerate}

The paper is organised as follows.
Section \ref{sec:background} describes the relevant background, Section \ref{sec:our_contribution} presents our contributions to knowledge, Section \ref{sec:solutions} shows the existing approaches that try to cope with the fork problem and, eventually, Section \ref{sec:conclusion} discuss the results and propose possible future directions.

\section{Background}\label{sec:background}

\subsection{Fuzz testing}

\begin{figure}[ht]
  \centering
  \begin{tikzpicture}[
    round/.style={circle, draw=black!100, fill=black!5, thick, minimum size=10mm},
    box/.style={rectangle, draw=black!100, fill=black!5, thick, minimum size=5mm, text width=15mm, align=center},
    parallelogram/.style={trapezium, draw,trapezium left angle=240, trapezium right angle=120, draw=black!100, fill=black!5, thick, minimum size=5mm},
    title/.style={font=\color{black!100}\ttfamily},
    criteria/.style={circle, draw=black!100, thin, minimum size=1mm},
    ] 
    
    \node[round](start){Start};
    \node[parallelogram](seeds)[below=0.8cm of start]{Seeds selection $^I$};
    \node[box](testcase)[below=of seeds]{Input generation};
    
    \node (exec) [ below=of testcase] {\makecell[l]{System execution } };
    
    \coordinate (prova) at ($(exec.east)+(-0.2cm,0cm)$);

    \node (time) [below=of exec, box] { Hangs detector };
    \node (anom) [left=of time, box] { Bugs detector };
    
    \node (cov) [right=of time, box] { Code coverage detector};

    \node (container) [draw=black!100, thick, fit={(anom) (cov) (time) (exec)}] {};

    \node[parallelogram](report)[below=of container]{Report $^O$};
    \node[round](end)[below=0.8cm of report]{End};
    
    \draw[->] (start) -- (seeds);
    \draw[->] (seeds) -- (testcase);
    \draw[->] (testcase) -- (container);
    \draw[->] (container) -- (report);
    \draw[->] (report) -- (end);
    \end{tikzpicture}
    \caption{Coverage-guided fuzzing process}
    \label{fig:coverage}
\end{figure}
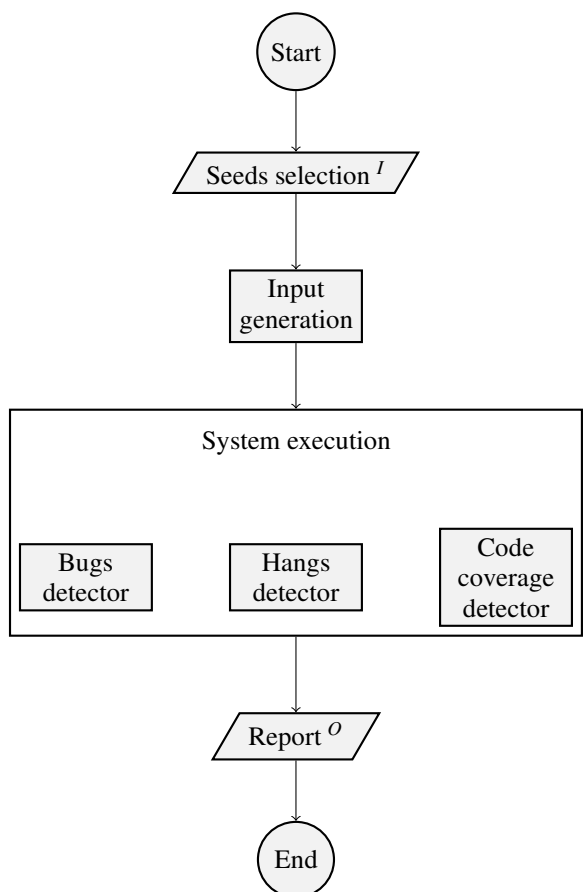

Fuzzing is an automated testing technique pioneered by Miller et al.\cite{miller1990empirical} in 1990 to test UNIX utilities.
As outlined in Figure \ref{fig:coverage}, coverage-guided fuzzing is composed at least of seed selection, input generation and system execution.
\par
\textit{1) Seeds selection.} The user must provide some input messages (seeds) representative of some usual inputs for the system. 
\par
\textit{2) Input generation.} The core of every fuzzer is the generation of slightly malformed input messages to forward to the software under test. 
A fuzzer is as efficient as the generated inputs are able to break the system. 
According to the approach used to generate the messages, the fuzzers may be classified into:
 \begin{itemize}
     \item \textit{dumb}: generate random strings (as the first fuzzer \cite{miller1995fuzz} did);
     \item \textit{dumb mutational}: blindly mutate seed messages provided by the user;
     \item \textit{grammar-based}: leverage the grammar of the system to craft the input messages;
     \item \textit{smart mutational} (often called \textit{evolutionary}): require a sample of inputs and leverage \textit{feedback mechanisms} to craft system-tailored messages. An example of feedback mechanisms is the code coverage feedback, explored in Section \ref{sec:coverage}.
 \end{itemize}
\par
\textit{3) System execution.} Each execution of the fuzzer involves three components:
\begin{itemize}
    \item \textit{Bugs detector}: it reports eventual bugs. The majority of the bugs detectors only report crashes, however for many systems, also a weird deviation from the happy flow of the protocol may represent significant security issues;
    \item \textit{Hangs detector}: it detects program execution hangs;
    \item \textit{Code coverage detector}: as further explained in Section \ref{sec:coverage}, the code coverage represents one of the feedbacks the fuzzer leverages to improve the quality of the input messages.
\end{itemize}

\subsection{Coverage-Guided Fuzzing}\label{sec:coverage}

\textit{Smart mutational} fuzzers use feedback mechanisms to steer the generation of the messages. 
Different types of feedback mechanisms exist \cite{shahid2011study}, and often different terms are used to express the same idea.
To avoid further noise, in this work we use the term \textit{code coverage} to express the lines of code that are reached by a specific message.

Code coverage fuzzers need to recompile the code with ad-hoc compilers (e.g. the AFL compiler) to instrument the code and obtain run-time information. 

AFL \cite{zalewski2017american}, for example, instruments the code to fill a bitmap that represents the lines of the code covered by the inputs.

Later, it uses this bitmap to assign a higher score to messages able to explore previously unseen lines of code.

\subsection{Inter-Process Communication}\label{sec:ipc}

Operating systems provide \textit{system calls} to perform different tasks (e.g. writing and reading files, accessing hardware services, creating and executing new processes).
On UNIX systems, new processes are created by using the \textit{fork system call}\cite{tanenbaum2009modern}.
In short, the first process, called \textit{parent process}, generates a clone, called \textit{child process}, that is an exact copy of the parent process.
After the fork, file descriptors and registers are duplicated, thus a change in one of the processes does not affect the other one. Also, the parent and child process will follow \textit{separate execution paths}.

\begin{table*}[t]\centering
\begin{tabular}{ |c|c|c|c|c|c|c|}
\hline
 Fuzzer &  Based on  & Monitor technique & \makecell[c]{Bugs\\Detection \\(C1)} & \makecell[c]{Hangs\\Detection \\(C2)} & \makecell[c]{ Code\\coverage \\ (C3)}   \\ 
 \hline
 AFL\cite{zalewski2017american} & - &  POSIX signals & \texttimes &  \texttimes &  \checkmark \\
 \hline
 AFL++\cite{fioraldi2020afl++} & AFL &  POSIX signals & \texttimes &  \texttimes &  \checkmark \\ 
 \hline
 AFLFast\cite{bohme2017coverage} & AFL &  POSIX signals & \texttimes &  \texttimes &  \checkmark \\  
 \hline
 AFLSmart\cite{pham2021smart} & AFL &  POSIX signals & \texttimes &  \texttimes &  \checkmark \\   
 \hline
 Eclipser\cite{choi2019grey} & AFL & POSIX signals & \texttimes &  \texttimes &  \checkmark \\   
 \hline
 FairFuzz\cite{lemieux2018fairfuzz} & AFL &  POSIX signals & \texttimes &  \texttimes &  \checkmark \\   
 \hline
 lafintel\cite{besler2016circumventing} & AFL &  POSIX signals & \texttimes &  \texttimes &  \checkmark \\   
 \hline
 AFLnwe\footnote{\url{https://github.com/thuanpv/aflnwe}} &  AFL & POSIX signals & \texttimes &  \texttimes &  \checkmark \\  
 \hline
  AFLNet\cite{pham2020aflnet} &  AFL &  POSIX signals & \texttimes &  \texttimes &  \checkmark \\ 
 \hline
 MOpt-AFL\cite{lyu2019mopt} & AFL &  POSIX signals & \texttimes &  \texttimes &  \checkmark \\   
 \hline
 StateAFL\cite{natella2021stateafl} &  \makecell[l]{- AFL\\- AFLNet} &  POSIX signals & \texttimes &  \texttimes &  \checkmark \\ 
  \hline
 LibFuzzer\footnote{\url{https://llvm.org/docs/LibFuzzer.html}} & - &  \makecell[l]{- UBSAN\\- ASAN\\- MSAN} & \checkmark &  \texttimes &  \checkmark \\   
 \hline
 Entropic\cite{bohme2020boosting} & LibFuzzer & \makecell[l]{- UBSAN\\- ASAN\\- MSAN} & \checkmark &  \texttimes &  \checkmark \\   
 \hline
 Honggfuzz\footnote{\url{https://honggfuzz.dev/}} & - & ptrace (Linux) & \checkmark &  \texttimes &  \checkmark \\   
 \hline
 \end{tabular}
\caption{Coverage guided fuzzers evaluation}
\label{table:fuzzers_evaluation}
\end{table*}

\section{Our contribution}\label{sec:our_contribution}

This paper aims to understand how the state-of-the-art coverage-guided fuzzers deal with software under tests containing forks. 

It was not obvious to come up with a way to compare and contrast the various tools. We devised a novel property, the \textit{fork awareness}, that must be satisfied when a fuzzer deals with forks effectively and efficiently.
As we shall see below, fork awareness rests upon three aspects representing the ability to deal with child processes.

Also, we evaluate the novel property over the most widely used fuzzers from two benchmark frameworks, reaching a total of 14 evaluated tools, 11 drawn from FuzzBench and 3 from ProFuzzBench.

\subsection{Fork-awareness}\label{sec:fork-awareness}
Abstractly, fork awareness insists that every fuzzer should address the child process as the parent one. 
During the system execution, the system monitor should detect bugs or hangs regardless of their location and the coverage should be measured also in the child process. This is formalised through Definition~\ref{def:1}.

\begin{definition}\label{def:1}
\emph{A coverage-guided fuzzer is} fork-aware \emph{if it can detect bugs and hangs and measure coverage in the same way for both the child and the parent's branch.}
\end{definition}
The three aspects in this definition are called:
\begin{enumerate}[label={\textbf{[C.\arabic*]}}]
  \item \textbf{Child bugs detection}: any anomaly is reported also if it occurs in child processes;
  \item \textbf{Child hangs detection}: any infinite hang is reported also if it occurs in child processes;
  \item \textbf{Child code coverage}: code coverage is measured also for child processes.
\end{enumerate}

\subsection{Example challenges}\label{sec:c}

We wrote three simple C programs to use as challenges for the fuzzers, namely to test whether the fuzzers satisfy the aspects given above.
\newline
\begin{enumerate}[label=\alph*)]
\item \textit{Bugs detection challenge}:
\begin{small}
\begin{verbatim}
1  if(fork()==0){ //Child process
2    raise(SIGSEGV); //Simulated crash
3  } else { //Parent process
4    wait(NULL); //Waiting child 
5                //termination
6  }
\end{verbatim}
\end{small}
The snippet sends a \textit{SIGSEGV} signal to simulate a bug in the child process.
This signal is used to report a segmentation fault, i.e. a memory access violation, which is common in programs written in low-level languages.
The fuzzer must detect this bug also after the parent's termination. 
\newline
\item \textit{Hangs detection challenge}:
\begin{small}
\begin{verbatim}
1  if(fork()==0){ //Child process
2    while(1){ ; } //Simulation of
3                  //blocking code
4  } 
\end{verbatim}
\end{small}
The snippet simulates an infinite loop in the child process.
The fuzzers must report processes still in execution after the loop and must kill child processes at the end of the fuzzing campaign, avoiding pending process executions.
\newline
\item \textit{Code coverage challenge}:
\begin{small}
\begin{verbatim}
1  pid_t pid = fork();
2  if(pid==0){ //Child process
3    if(data %2 == 0){ do_something(); } 
4       else { do_something(); }
5    if(data %3 == 0){ do_something(); } 
6       else { do_something(); }
7    if(data %5 == 0){ do_something(); } 
8       else { do_something(); }
9    if(data %7 == 0){ do_something(); } 
10      else { do_something(); }
11    } 
12 else { //Parent process
13      wait(NULL); //Waiting child 
14                //termination
15      }
\end{verbatim}
\end{small}
\end{enumerate}
This snippet simulates a child with several branches.
%
A fuzzer must cover and consider every child's branches.
\newline


We run the 14 fuzzers over these challenges and organised the results in Table~\ref{table:fuzzers_evaluation}. We noticed that none of the fuzzers succeeded through all three challenges.
\subsection{Testbed}
We decided to analyse only the coverage-guided fuzzers present in FuzzBench \cite{metzman2021fuzzbench} and ProFuzzBench \cite{natella2021profuzzbench} even though the property applies to every coverage-guided fuzzer.
All fuzzers were executed on an \textit{Ubuntu 20.04} server machine and all our source codes are freely available online\footnote{\url{https://github.com/marcellomaugeri/forks-break-afl}} so that our experiments are fully reproducible.

\subsection{Fuzzers evaluation}
We run all selected fuzzers against our three example challenges.
Table~\ref{table:fuzzers_evaluation} summarises our findings.

All the fuzzers based on AFL use \textit{POSIX signals} and a bitmap respectively to report bugs and keep track of the code coverage.

As shown in the Table~\ref{table:fuzzers_evaluation}, while the bitmaps are able to keep track of the child's code coverage, bugs triggered in the child's processes are not detected since AFL catches signals from the main process only, as pointed out in the documentation\footnote{\url{https://github.com/google/AFL/blob/master/README.md}}.
The only fuzzers able to detect bugs in the child process are LibFuzzer\footnote{\url{https://llvm.org/docs/LibFuzzer.html}}, Entropic\cite{bohme2020boosting} and Hongfuzz\footnote{\url{https://honggfuzz.dev/}}, as discussed in more detail below:
\begin{itemize}
    \item \textit{LibFuzzer} \footnote{\url{https://llvm.org/docs/LibFuzzer.html}} and \textit{Entropic}\cite{bohme2020boosting} employ a set of sanitizers\footnote{AddressSanitizer, UndefinedBehaviorSanitizer and MemorySanitizer} to report bugs. 
These mechanisms make the fuzzers able to find the bug in Challenge 1 and measure the different code paths in Challenge 3, thereby satisfying challenges C.1 and C.3, as seen above.
Unfortunately, challenge C.2 is not satisfied since the fuzzer cannot detect hangs in the child process.
\item \textit{Honggfuzz} supports different software/hardware feedback mechanisms and a low-level interface to monitor targets.
When executed on Linux machines, Honggfuzz uses the \textit{ptrace} system call to manage processes. This mechanism allows the fuzzer to capture a wide range of signals.
As shown in Table~\ref{table:fuzzers_evaluation}, the use of ptrace (along with the \textit{SanitizerCoverage}) allows the fuzzer to detect bugs and to consider coverage also in the child process.
Unfortunately, neither this mechanism is able to detect hangs in the child process.
\end{itemize}

In summary, while all selected fuzzers detect the code coverage (C3), none detect hangs (C2) and only a few detect bugs (C1) in the child process.
The evaluation underlines that:
\begin{itemize}
    \item \textit{Loops detection challenge} is the most difficult because fuzzers do not wait for all the child processes but only for the main one;
    \item \textit{Code coverage challenge} is the easiest because the instrumentation allows measuring coverage from the execution, regardless of the process involved;
    \item \textit{Bug detection challenge} depends on the technique used to observe bugs, as well as the use of sanitisers.
\end{itemize}
We interpret this general outcome as a clear call for future research and developments.

\section{Existing solutions}\label{sec:solutions}

Nowadays the only solutions to fuzz programs that use forks are manually modifying the code or breaking the multi-process nature of the system (by employing tools like defork\footnote{\url{https://github.com/zardus/preeny/blob/master/src/defork.c}}) in order to get rid of the forks.

Unfortunately, making modifications to the code, as pointed out in the AFLNet documentation \footnote{\url{https://github.com/aflnet/aflnet}}, to remove all the forks is a challenging and error-prone task and break the multi-process nature of the system often leads to weird system behaviours.
The only solution, therefore, remains to modify the fuzzers.

\section{Conclusions}\label{sec:conclusion}

This paper analyses the fork awareness of the coverage-guided fuzzers using three different aspects. 
The analysis conducted on 14 well-known fuzzers highlights that while is it clear how important is to handle multi-process programs, the majority of the fuzzers overlook the problem.
11 of 14 fuzzers are not able to detect bugs in the child process.
The intuition behind these outcomes is related to the way these fuzzers detect bugs. All the AFL-derived fuzzers use signals (SIGSEGV, SIGABRT, etc) to detect bugs and this mechanism misses bugs in child processes.
We noticed that dealing with forks is not the only problem and other issues may be related to the IPC scheduling.
For example, the IPC may influence the success of the fuzzing process since some bugs may be triggered only after a specific process schedule and only after access to a particular cell of memory.
We believe this paper represents a first step towards the devising of fuzzers aware of the eventual multiprocess nature of the software.
The first step to achieve this goal might be the implementation of a \textit{loop detector} at an early stage, e.g. by leveraging a dynamic library to keep track of all process identifiers of forked processes.
To summarise, this work not only provides the first concrete way to evaluate the fuzzers according to their fork awareness but sheds light for the first time on a class of problems that have been ignored until now, showing interesting future directions.

\bibliographystyle{apalike}

\bibliography{bibliography}

\begin{thebibliography}{}

\bibitem[Besler and Frederic, 2016]{besler2016circumventing}
Besler and Frederic (2016).
\newblock Circumventing fuzzing roadblocks with compiler transformations.
\newblock \url{https://lafintel.wordpress.com}.

\bibitem[Boehme et~al., 2021]{boehme2021fuzzing}
Boehme, M., Cadar, C., and Roychoudhury, A. (2021).
\newblock Fuzzing: Challenges and reflections.
\newblock {\em IEEE Softw.}, 38(3):79--86.

\bibitem[Bohme et~al., 2020]{bohme2020boosting}
Bohme, M., Manes, V.~J., and Cha, S.~K. (2020).
\newblock Boosting fuzzer efficiency: An information theoretic perspective.
\newblock In {\em Proceedings of the 28th ACM Joint Meeting on European
  Software Engineering Conference and Symposium on the Foundations of Software
  Engineering}, pages 678--689.

\bibitem[Bohme et~al., 2017]{bohme2017coverage}
Bohme, M., Pham, V.-T., and Roychoudhury, A. (2017).
\newblock Coverage-based greybox fuzzing as markov chain.
\newblock {\em IEEE Transactions on Software Engineering}, 45(5):489--506.

\bibitem[Choi et~al., 2019]{choi2019grey}
Choi, J., Jang, J., Han, C., and Cha, S.~K. (2019).
\newblock Grey-box concolic testing on binary code.
\newblock In {\em 2019 IEEE/ACM 41st International Conference on Software
  Engineering (ICSE)}, pages 736--747. IEEE.

\bibitem[Fioraldi et~al., 2020]{fioraldi2020afl++}
Fioraldi, A., Maier, D., Ei{\ss}feldt, H., and Heuse, M. (2020).
\newblock Afl++: Combining incremental steps of fuzzing research.
\newblock In {\em 14th USENIX Workshop on Offensive Technologies (WOOT 20)}.

\bibitem[Hazimeh et~al., 2020]{hazimeh2020magma}
Hazimeh, A., Herrera, A., and Payer, M. (2020).
\newblock Magma: A ground-truth fuzzing benchmark.
\newblock {\em Proceedings of the ACM on Measurement and Analysis of Computing
  Systems}, 4(3):1--29.

\bibitem[Lemieux and Sen, 2018]{lemieux2018fairfuzz}
Lemieux, C. and Sen, K. (2018).
\newblock Fairfuzz: A targeted mutation strategy for increasing greybox fuzz
  testing coverage.
\newblock In {\em Proceedings of the 33rd ACM/IEEE International Conference on
  Automated Software Engineering}, pages 475--485.

\bibitem[Lyu et~al., 2019]{lyu2019mopt}
Lyu, C., Ji, S., Zhang, C., Li, Y., Lee, W.-H., Song, Y., and Beyah, R. (2019).
\newblock $\{$MOPT$\}$: Optimized mutation scheduling for fuzzers.
\newblock In {\em 28th USENIX Security Symposium (USENIX Security 19)}, pages
  1949--1966.

\bibitem[Metzman et~al., 2021]{metzman2021fuzzbench}
Metzman, J., Szekeres, L., Maurice Romain~Simon, L., Trevelin~Sprabery, R., and
  Arya, A. (2021).
\newblock {FuzzBench: An Open Fuzzer Benchmarking Platform and Service}.
\newblock In {\em Proceedings of the 29th ACM Joint Meeting on European
  Software Engineering Conference and Symposium on the Foundations of Software
  Engineering}, ESEC/FSE 2021, pages 1393--1403, New York, NY, USA. Association
  for Computing Machinery.

\bibitem[Miller et~al., 1990]{miller1990empirical}
Miller, B.~P., Fredriksen, L., and So, B. (1990).
\newblock An empirical study of the reliability of unix utilities.
\newblock {\em Communications of the ACM}, 33(12):32--44.

\bibitem[Miller et~al., 1995]{miller1995fuzz}
Miller, B.~P., Koski, D., Lee, C.~P., Maganty, V., Murthy, R., Natarajan, A.,
  and Steidl, J. (1995).
\newblock Fuzz revisited: A re-examination of the reliability of unix utilities
  and services.
\newblock Technical report, University of Wisconsin-Madison Department of
  Computer Sciences.

\bibitem[Natella, 2022]{natella2021stateafl}
Natella, R. (2022).
\newblock Stateafl: Greybox fuzzing for stateful network servers.
\newblock {\em Empirical Software Engineering}, 27(7):191.

\bibitem[Natella and Pham, 2021]{natella2021profuzzbench}
Natella, R. and Pham, V.~T. (2021).
\newblock Profuzzbench: A benchmark for stateful protocol fuzzing.
\newblock In {\em ISSTA 2021 - Proceedings of the 30th ACM SIGSOFT
  International Symposium on Software Testing and Analysis}, pages 662--665.
  Association for Computing Machinery, Inc.

\bibitem[Pham et~al., 2020]{pham2020aflnet}
Pham, V.-T., Bohme, M., and Roychoudhury, A. (2020).
\newblock Aflnet: a greybox fuzzer for network protocols.
\newblock In {\em 2020 IEEE 13th International Conference on Software Testing,
  Validation and Verification (ICST)}, pages 460--465. IEEE.

\bibitem[Pham et~al., 2021]{pham2021smart}
Pham, V.-T., Böhme, M., Santosa, A.~E., Căciulescu, A.~R., and Roychoudhury,
  A. (2021).
\newblock Smart greybox fuzzing.
\newblock {\em IEEE Transactions on Software Engineering}, 47(9):1980--1997.

\bibitem[Shahid et~al., 2011]{shahid2011study}
Shahid, M., Ibrahim, S., and Mahrin, M.~N. (2011).
\newblock A study on test coverage in software testing.
\newblock {\em Advanced Informatics School (AIS), Universiti Teknologi
  Malaysia, International Campus, Jalan Semarak, Kuala Lumpur, Malaysia}.

\bibitem[Tanenbaum, 2009]{tanenbaum2009modern}
Tanenbaum, A. (2009).
\newblock {\em Modern operating systems}.
\newblock Pearson Education, Inc.,.

\bibitem[Zalewski, 2017]{zalewski2017american}
Zalewski, M. (2017).
\newblock American fuzzy lop.
\newblock \url{https://lcamtuf.coredump.cx/afl/}.

\end{thebibliography}

\end{document}